# BAYESIAN MATCHING FOR X-RAY AND INFRARED SOURCES IN THE MYSTIX PROJECT

Tim Naylor[*,1], Patrick S. Broos[2], Eric D. Feigelson[2]

*Draft version September 17, 2013*

## ABSTRACT

Identifying the infrared counterparts of X-ray sources in Galactic Plane fields such as those of the MYStIX project presents particular difficulties due to the high density of infrared sources. This high stellar density makes it inevitable that a large fraction of X-ray positions will have a faint field star close to them, which standard matching techniques may incorrectly take to be the counterpart. Instead we use the infrared data to create a model of both the field star and counterpart magnitude distributions, which we then combine with a Bayesian technique to yield a probability that any star is the counterpart of an X-ray source. In our more crowded fields, between 10 and 20% of counterparts that would be identified on the grounds of being the closest star to X-ray position within a 99% confidence error circle are instead identified by the Bayesian technique as field stars. These stars are preferentially concentrated at faint magnitudes. Equally importantly the technique also gives a probability that the true counterpart to the X-ray source falls beneath the magnitude limit of the infrared catalog. In deriving our method, we place it in the context of other procedures for matching astronomical catalogs.

*Keywords:* methods: data analysis — methods:statistical — stars: pre-main sequence — stars: formation — infrared: stars — X-rays: stars

## 1. INTRODUCTION

A central goal of the Massive Young Star-Forming Complex Study in Infrared and X-rays (MYStIX; Feigelson et al. 2013) is to obtain a rich and high-quality census of stars belonging to complex star formation regions within 4 kpc of the Sun. A first step is to obtain catalogs of X-ray, near-infrared and mid-infrared sources from single waveband surveys (see Kuhn et al. 2013a; Townsley & Broos 2013; King et al. 2013; Kuhn et al. 2013b). The X-ray selection is effective in discriminating young stars from older Galactic field stars, while the infrared photometry is needed to characterize the properties of the young stars: luminosity, surface temperature, absorption, and infrared-excess from a circumstellar dusty disk. A crucial step in our analysis, therefore, is to reliably cross-identify X-ray sources with infrared sources, which is an example of a fundamental problem in astronomy, that of matching one catalog of sources to another.

The very simplest cases of catalog matching are "obvious" in the sense that if one overlays the two catalogs in sky coordinates, correct pairings lie close to each other compared with the typical separations between stars internally within either catalog. The problem becomes more difficult when the area covered by the uncertainty in the position in one catalog (which we shall call the master) approaches or exceeds the reciprocal of the density of sources in the other catalog (which we will term the slave). One can then imagine drawing circles in the slave catalog (with a radius derived from the uncertainty in position) that encompass likely counterparts. The key question is then which of the possible counterparts to choose. The simplest answer is to take the star closest to the X-ray position, which we shall term proximity-only matching.

Such proximity-based identification criteria will give intuitively incorrect answers as survey improvements produce deeper slave catalogs with higher source densities. For example, matching the master catalog against a relatively shallow slave catalog may produce the correct, relatively bright counterpart. However, as one produces ever deeper slave catalogs, it is inevitable that a faint star will be detected closer to the X-ray position than the true counterpart. Proximity-only matching would then, incorrectly, identify the faint star as the counterpart.

This displacement problem is a significant issue for the crowded fields in the MYStIX sample (see Figure 6 in Feigelson et al. 2013). For example, in the Trifid Nebula field at Galactic longitude $7°$ where the Galactic field star density is extremely high, we have detected approximately 600 faint X-ray sources from the *Chandra* X-ray Observatory, 26 000 *Spitzer* Space Telescope mid-infrared (MIR) sources and 76 000 sources from the UK InfraRed Telescope (UKIRT) near-infrared (NIR) data. Obviously the great majority of these infrared stars are irrelevant to the X-ray sources and the targeted star forming region, but tied to this is a more subtle difference about the depth reached by each survey. In uncrowded fields the UKIRT data reach down to $K = 17.5$ at a signal-to-noise of 10, but there are very few reliable counterparts to X-ray sources fainter than $K = 15$.[4] Conversely the *Spitzer* data are probably well matched in depth to the X-ray data, as there are significant numbers of counterparts to X-ray sources at the limiting magnitude of the MIR catalogs, but there are normally not many more counterparts in the near-infrared catalog, suggesting that the MIR sample is fairly complete.

[*] timn@astro.ex.ac.uk
[1] School of Physics, University of Exeter, Stocker Road, Exeter EX4 4QL, UK
[2] Department of Astronomy & Astrophysics, Pennsylvania State University, 525 Davey Lab, University Park, PA 16802

---

[4] This arises form a well-known correlation between X-ray and photospheric luminosity in pre-main sequence stars (Telleschi et al. 2007).

The crucial point is that for the UKIRT data, not only are most stars irrelevant to the X-ray sources, but the counterparts lie preferentially amongst the brighter IR (infrared) objects. Clearly the key to solving the displacement problem must be to use information about the magnitude distribution of the counterparts compared with that of the field stars. Specifically, if a very faint star is found close to the X-ray position in the UKIRT data, it should not be taken as a likely counterpart, because such faint counterparts are very rare, but faint field stars are common.

There are other challenges the matching process presents. Although the positional accuracies of the UKIRT and *Spitzer* images are approximately constant across their respective fields-of-view, we must correctly treat the heteroscedasticity of the X-ray positional measurement errors arising from the differing counts in each source and the spatially varying *Chandra* point spread function. For bright, on-axis sources we find a mean offset between X-ray and NIR counterparts of around $0.1''$, which is probably dominated by the astrometric accuracy of the NIR data. Towards the periphery of the *ACIS* field this can decline to several arcseconds, which can lead to two or more possible infrared counterparts. Furthermore, the infrared background source density can be spatially highly variable due to molecular cloud obscuration and H II region nebular emission. So it is clear the criteria for counterpart identification must adapt to the quality of the detection in the X-ray data, as well as its environment in the UKIRT and *Spitzer* images. Finally, an unknown fraction of the X-ray sources are extra-galactic interlopers without any detectable infrared counterpart; such sources should be marked as un-matched, rather than identified with a faint star far from the X-ray position.

A potential problem is that we are not simply interested in pairing IR sources with X-ray sources, but also in the spectral energy distribution of a source over all three bands (Povich et al. 2013). This requires asking whether a single object position is consistent with the positions (and their uncertainties) for the sources observed in each waveband. Budavári & Szalay (2008) and Storkey et al. (2005) consider this problem in some detail, but practically it is only an issue when there are several catalogs with large uncertainties in position. In our case, the X-ray catalog normally has the largest uncertainties, with the MIR and NIR data both having smaller (and similar) positional accuracies, although the WFCAM camera on UKIRT can resolve a crowded complex with much better resolution than *Spitzer*.

We therefore adopt the simplifying strategy of matching the NIR and MIR catalogs as slaves to the X-ray catalog, and only if the MIR and NIR catalog positions differ might one wish to consider the case that there may be two sources present. Practically, however, the resolution and sensitivity of the UKIRT and *Spitzer* catalogs are well matched, and they differ relatively little in wavelength, and so we believe such cases will be extremely rare. This is supported by the low error rate for proximity-only matching between the two catalogs (Povich et al. 2013). We will return to the problem of multiple catalog matching in Section 8

Within the MYStIX data reduction pipeline (see Figure 3 in Feigelson et al. 2013) matching lies between the creation of the X-ray and IR catalogs (Kuhn et al. 2013a; Townsley & Broos 2013; King et al. 2013; Kuhn et al. 2013b), and classification with the naive Bayes classifier (Broos et al. 2013). It is worth emphasizing that there is a clear split between the observational information used in the matching process (the observed magnitude distributions of the IR counterparts and field stars) and the astrophysical understanding (colors and magnitudes for particular types classes of object) used in the Bayes classifier.

## 2. THE OUTLINE SOLUTION

Section 1 laid out the fundamental problem faced in identifying the infrared counterparts to the X-ray sources when the true counterparts have a magnitude distribution systematically brighter than the Galactic field star magnitude distribution. With just positional information, a faint field star close to the X-ray position appears a better candidate than the true counterpart which may be a slightly more distant but much brighter star. To solve this problem requires using information about the magnitude distribution of both the IR catalog as a whole and the IR magnitudes of the counterparts themselves.

There are two obvious sources for such information, either an astrophysical understanding of the sources, or a study of the properties of the X-ray counterparts as a whole. As stated in Section 1, we will very deliberately avoid using astrophysical information, as this is the domain of the naive Bayes classification which follows our matching process, and is described in Broos et al. (2013). It would be incorrect, for example, use a training set such as the Chandra Orion Ultradeep Survey (Feigelson et al. 2005) to predict the magnitude of our counterparts, since this would be explicitly including astrophysical understanding. That leaves us with the possibility of using the data themselves to yield statistical information about the magnitude distributions of both the field stars and the counterparts. Dealing with the field stars is straightforward, an area around the X-ray position can be chosen, and a histogram of the magnitude distribution of the stars within this area can be created. For the counterparts we can use the distribution of the magnitudes of stars within the error circles, and subtract from it the expected field star distribution.

Such an approach fits within a broad theme in the literature of using not only the counterpart's celestial position, but some property such as its color, morphology, or spectral energy distribution (e.g. Roseboom et al. 2009), or even a combination of several properties (Rohde et al. 2005, 2006) to assign the likelihood of it being the counterpart. Although we do not know of such techniques being used for Galactic astronomy before, they are widely used in extra-galactic work, and most techniques trace their origins back to an increasingly influential paper by Sutherland & Saunders (1992). Given the crowded fields of the MYStIX project, it is clear that such techniques have much to offer for work in the Galactic plane, though as discussed in Section 4 that same crowding requires special measures to determine the counterpart magnitude distribution.

In what follows, the likelihood that any given star is the counterpart of a given X-ray source is derived, given the magnitude distributions of both the field and coun-terpart stars (Section 3 or Appendix A). This leads to a similar result to that of Sutherland & Saunders (1992), but extended to (i) yield the probability that the X-ray source has no counterpart in the IR catalog, and (ii) show that a slightly different definition of the magnitude distribution functions is more appropriate. We then derive a method to determine the counterpart magnitude distribution which is appropriate for Galactic fields that are not only crowded, but also have a stellar density which can change rapidly with position on the sky (Section 4). Finally, Section 5 deals with the practical issues of applying the theory, using the Trifid Nebula field as an example. Table 1 gives definitions of the mathematical symbols used throughout the paper.

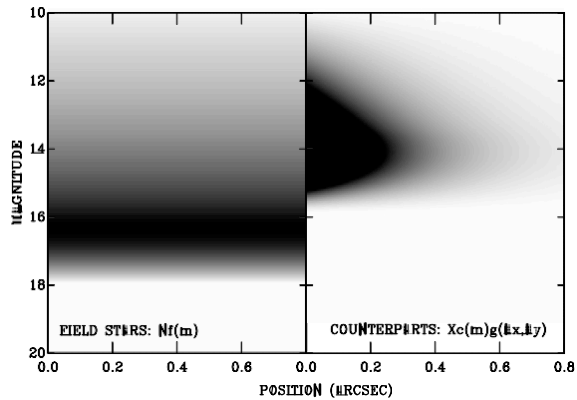

**Figure 1.** Probability density functions for field stars and counterparts as functions of magnitude and position. The grey scale represents values illustrative of the general form of these functions, not real data.

### 3. THE FUNDAMENTAL EQUATION

We wish to derive the likelihood that a given star is a counterpart, or that the counterpart is not in the catalog. In this Section we will derive equations for these quantities using a route which drives an understanding of the problem by beginning with the densities of stars and counterparts. An alternative derivation is given in Appendix A, based around Bayes' Rule. This exposes both some implicit assumptions and the Bayesian roots of the method.

Consider the density of stars as a function of distance from the X-ray position and magnitude. This is a three-dimensional volume, but in Figure 1 it is illustrated using radial distance from the X-ray position so there is just one position coordinate. The value of the probability density function is represented as the density of points in the plane, and this function is shown for both field stars and counterparts. The density of field stars in this example is independent of position, but increases with magnitude until it reaches the limiting magnitude of the catalog at around $m = 18$. Conversely, the true counterparts tend to be brighter than the field stars, and their density declines with distance from the X-ray position.

To express Figure 1 mathematically requires the probability density function representing the distribution of field stars with magnitude, evaluated at a magnitude $m$, which is $f(m)$. Thus, given that there is a field star, $f(m)dm$ is the probability that its magnitude lies between $m$ and $m + dm$, and so $f(m)$ has units of mag$^{-1}$, and integrates to one[5]. The left-hand panel of Figure 1 represents the density of field stars per unit area on the sky, per magnitude, and so if the total number of field stars per unit area in our catalog is $N$, then the probability density at any point in our volume is $Nf(m)$, which has units of mag$^{-1}$ arcsec$^{-2}$. In contrast the sky density of counterparts is distributed according to the normalized two-dimensional Gaussian function that represents the uncertainty ellipse of the X-ray position, which we label $g(\Delta x, \Delta y)$ where the coordinates are with respect to the X-ray position. The probability distribution for the magnitudes of the counterparts is given by $Xc(m)$, where $X$ is the fraction of X-ray objects that have counterparts in the catalog, and $c(m)$ is the probability density function in magnitude. Thus the density of X-ray counterparts in our volume is $Xc(m) g(\Delta x, \Delta y)$, again with units of mag$^{-1}$ arcsec$^{-2}$.

It is illuminating to draw the distinction between quantities such as $Xc(m)$ and $c(m)$. $f(m)$ and $c(m)$ are true probability density functions which are normalized to integrate one, and answer the question "given a star, what is the probability that its magnitude lies between $m$ and $m + dm$". In contrast $Xc(m)dm$ answers the question "what is the probability that the counterpart to a given star lies between magnitudes $m$ and $m + dm$", with the extra term covering the possibility that there is no counterpart in our IR catalog. Functions such as $Xc(m)$ are not normalized to integrate one, and so we shall refer to them simply as a probability functions, not probability density functions.

It is now clear that the probability of finding a field star at a magnitude $m_j$ is proportional to $Nf(m_j)$. By extension the likelihood of finding an IR source $i$ which is the counterpart, at position $(\Delta x, \Delta y)$ with magnitude $m_i$, **and** an IR source $j$ which is a field star with magnitude $m_j$ is given by

$$L(i, j) \propto Xc(m_i) g(\Delta x, \Delta y) \times Nf(m_j). \quad (1)$$

With $m_i$ and $m_j$ frozen at measured values, Equation 1 is a function of the indices $i$ and $j$ that define the two-source hypothesis stated above. Equation 1 is formally known as a "likelihood function"; the function arguments are the parameters of the hypothesis, $i$ and $j$. If we now consider all stars in our IR catalogue, the probability that star $i$ is the counterpart and all the other stars are field stars is given by

$$P(i) = KXc(m_i) g(\Delta x, \Delta y) \prod_{i,j=j}^{n} Nf(m_j) \quad (2)$$

$$= K \frac{Xc(m_i) g(\Delta x, \Delta y) \prod_j Nf(m_j)}{Nf(m_i)}, \quad (3)$$

where $K$ is a normalization constant.

The probability that no star is the counterpart, is simply the likelihood that every star is a field star multiplied by the likelihood that the counterpart is not in the X-ray

---

[5] Strictly $f(m)$ is also a function of position in the IR catalog, and as described in Section 5 our implementation requires that changes on the scale of tens of error circle radii must be small. Hence f(m) should be taken to mean its value close to the X-ray position, though not sufficiently close that crowding by counterparts of X-ray sources affects it.

**Table 1**
A summary of the meanings of the symbols used.

| Symbol | Description |
|---|---|
| $A$ | The area of an error circle. |
| $b(m)\mathrm{d}m$ | The probability that the brightest star in a given error circle (be it counterpart or field) has a magnitude between $m$ and $m+\mathrm{d}m$. |
| $B(m^t)$ | The integral of the above from $m = -\infty$ to $m^t$, i.e. the probability that the brightest star is brighter than magnitude $m^t$. |
| $b_\mathrm{f}(m)\mathrm{d}m$ | The probability that the brightest field star in a given error circle has a magnitude between $m$ and $m+\mathrm{d}m$. |
| $B_\mathrm{f}(m^t)$ | The integral of the above from $m = -\infty$ to $m^t$, i.e. the probability that the brightest field star is brighter than magnitude $m^t$. |
| $c(m)\mathrm{d}m$ | The fraction of IR counterparts that have magnitudes between $m$ and $m+\mathrm{d}m$. |
| $C(m^t)$ | The integral of the above from $m = -\infty$ to $m^t$, i.e. the probability that an IR counterpart is brighter than magnitude $m^t$. |
| $D$ | The observed data catalog. |
| $f(m)\mathrm{d}m$ | The fraction of field stars with magnitudes between $m$ and $m+\mathrm{d}m$ in the region of the IR catalog close to the X-ray position, in the absence of any crowding effects due to counterparts. |
| $F(m^t)$ | The integral of the above from $m = -\infty$ to $m^t$, i.e. the probability that a field star is brighter than magnitude $m^t$. |
| $g(\Delta x, \Delta y)\mathrm{d}x\mathrm{d}y$ | The fraction of IR counterparts that lie in box defined by $\Delta x$ to $\Delta x+\mathrm{d}x$ and $\Delta y$ to $\Delta y+\mathrm{d}y$ with respect to the X-ray position. |
| $H_0$ | The hypothesis that the IR counterpart is not in the catalog |
| $\tilde{H}_0$ | The hypothesis that the IR counterpart is in the catalog |
| $K$ | A normalization constant. |
| $L(i,j)$ | The likelihood that star $i$ is the counterpart and star $j$ and field star. |
| $M$ | The number of pixels occupied by stars. |
| $N$ | The total number of field stars per unit area in our IR catalog. |
| $O$ | The probability that an infinitesimal pixel on the sky is occupied by a star. |
| $P(0)$ | The probability that none of the stars in the catalog are the counterpart. |
| $P(i)$ | The probability that star $i$ is the IR counterpart of the X-ray source. |
| $T$ | The total number of pixels in the area of sky observed. |
| $X$ | The fraction of X-ray sources that have counterparts in the IR catalog. |
| $Z$ | The fraction of error circles that have a star (be it counterpart or field) within them that is in the IR catalog. |
| $Z_\mathrm{c}$ | The fraction of X-ray sources that have counterparts within a given error circle in the IR catalog. |
| $Z_\mathrm{f}$ | The fraction of error circles that have a field star within them that is in the IR catalog. |

catalog, and thus

$$P(0) = K(1-X)\prod_j^{TI} Nf(m_j). \quad (4)$$

All possible models for the distribution of stars in our position/magnitude volume have now been enumerated, and so $K$ can be determined since

$$P(0) + \sum P(i) = 1. \quad (5)$$

Substituting into Equations 3, 4 and 5 then yields,

$$P(0) = \frac{1-X}{1-X + \sum_j \frac{Xc(m_j)g(\Delta x_j, \Delta y_j)}{Nf(m_j)}} \quad (6)$$

and

$$P(i) = \frac{\frac{Xc(m_i)g(\Delta x_i, \Delta y_i)}{Nf(m_i)}}{1-X + \sum_j \frac{Xc(m_j)g(\Delta x_j, \Delta y_j)}{Nf(m_j)}} \quad (7)$$

Equation 7 is similar to Equation 5 of Sutherland & Saunders (1992), though there is a significant difference, which our derivation makes clear. Sutherland & Saunders (1992) use the intrinsic magnitude distributions which extend down to infinitely faint stars for the field and counterpart populations (called $n(m)$ and $q(m)$ respectively in their paper). Our $Nf(m)$ and $Xc(m)$ are the magnitude distributions in the catalogs, and so differ from $n(m)$ and $q(m)$ by the (magnitude dependent) completeness function for the catalog. The mathematics does not make this distinction clear since $c$ and $f$ only ever appear as a ratio, and so the completeness function cancels out. However, this distinction will become important in Section 4 since it is $c$ and $f$ that are determined from the catalog. To overcome this problem Sutherland & Saunders (1992) introduced a magnitude cut-off brighter than which the catalog was complete, below which no stars were detected. Our interpretation allows us to generalise their result to a completeness function that changes slowly with magnitude.

It is useful to compare these equations with other related techniques in common usage. There are two distinct threads in the literature that claim descendence from Sutherland & Saunders (1992). One decides how likely a star is to be a counterpart using only Equation (1) of Sutherland & Saunders (1992), the equivalent of $c(m) g(\Delta x, \Delta y)/Nf(m_j)$, which is often termed the likelihood ratio (e.g. Mann et al. 1997). They then use the distribution of likelihood ratios to determine how probable it is that a given star is the counterpart for a given X-ray source (e.g. Oyabu et al. 2005). This clearly does not take into account the other stars close to the error circle. Consider, for example, two X-ray positions. X-ray position one has a star at magnitude $m$ and distance $r$ from it, whilst X-ray position two has two stars of magnitude $m$ at distance $r$. Given that only one star per X-ray position can be the counterpart, the star close to X-ray position one should have a higher probability of being a counterpart than either of the stars close to position two; a result which the likelihood ratio fails to produce, but Equation 7 achieves.

The second thread related to Sutherland & Saunders (1992) does use the equivalent of Equation 7, which is sometimes called the reliability of the source[6]. This is exemplified by Fleuren et al. (2012). However these techniques do not explicitly calculate the probability that the counterpart of an individual X-ray source is not in the catalog ($P(0)$ as opposed to the global variable $X$), although Rutledge et al. (2000) attempts to do so using likelihood ratios. We believe the likelihood ratio alone is not a useful statistic.

---

[6] Confusingly, some authors who use only the likelihood ratio

## 4. THE COUNTERPART MAGNITUDE DISTRIBUTION

To use Equations 6 and 7, one needs to measure the magnitude distributions for both the field stars and the counterparts. Whilst the former is straightforward (it can be measured from areas of sky clear of X-ray sources), the latter is problematic. Previous work using the current approach to counterpart finding has determined the magnitude distribution for the counterparts by first measuring the magnitude distribution of all the stars within, say, a 68% confidence error circle. A distribution of field-star magnitudes is then subtracted from this, which is created from areas of sky where no X-ray sources are expected (e.g. Luo et al. 2010).

The problem in our application in the crowded Galactic Plane is that our density of counterparts within the error circles (∼ 1 per square arcsecond) is so high that they will tend to crowd out the field stars. For example, in some areas of our survey we expect there will be roughly one field star of around $K = 17$ in every two error circles. In a significant fraction of those error circles a bright (perhaps $K = 14$) true counterpart star will produce wings in the image that prevent detection of nearby faint field stars. Hence within such error circles the observations tend to be less sensitive and thus the catalog more incomplete, with a brighter limiting magnitude than in the full field. As a result the simple procedure such as that outlined by Luo et al. (2010) would yield a field-subtracted magnitude distribution for the counterparts which is negative at $K = 17$.

The impact this crowding effect can have on the derived magnitude distribution is not small, as illustrated in Figure 2. Here "counterparts" were inserted into the Trifid Nebula field according to the distribution in magnitude and the fraction of sources with X-ray counterparts found in Section 5 (the stars in Figure 2). Any fainter field stars within the error circles were then removed (as the error circles are typically less than an arcsecond in radius this is a good simulation of the crowding), unless there was a brighter field star within $0.8''$ of our simulated counterpart, in which case the counterpart was removed. We then selected all stars within typical 68% confidence circles of these stars, and subtracted from their magnitude distribution an estimated field-star magnitude distribution from a annular regions around each inserted star. The resulting estimate of the counterpart magnitude distribution (the dotted histogram of Figure 2) falls significantly short of the true distribution at faint magnitudes by, for example, a factor two at $K = 17$.

This over-subtraction problem is clearly present in the data of Mann et al. (1997), where it was thought to be simply noise, and was correctly identified by Rutledge et al. (2000). Of course, our problem is more extreme than those quoted above, because the MYStIX targets call the probability they derive from the distribution of likelihood ratios the reliability as well.

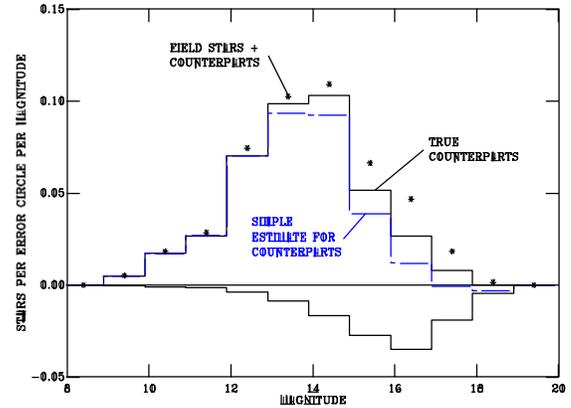

**Figure 2.** An example of the effects of using simple field star subtraction on the estimation of the counterpart magnitude distribution function for the Trifid Nebula field. The asterisks show the distribution of magnitudes for all stars in the error circles. The negative-going histogram is the negative of the distribution of field star magnitudes taken from annular areas around the error circles. The result of subtracting these two is the dashed blue histogram, which is an estimate of the counterpart magnitude distribution. This should be compared with the actual distribution of counterpart magnitudes which is the solid histogram. Note how this procedure fails to reproduce the input counterpart magnitude distribution, especially at faint magnitudes.

are crowded Galactic fields, as opposed to the relatively uncrowded extra-galactic case.

One possible way of correctly measuring the counterpart magnitude distribution is the proposed by Brusa et al. (2007). They chose a set of field stars whose magnitude distribution matched that of the X-ray sources, and then used the other stars that fell within a given radius of them to estimate the field star distribution which would be observed in the real error circles. Unfortunately our field star distribution changes rapidly with position, so we require an estimate close to the X-ray position.

To solve the problem requires a measurable quantity that is related to the counterpart magnitude distribution, but which is independent of the crowding. The distribution of counterpart magnitudes required has already been defined in Section 3, as the probability density function $c(m)$. We will represent cumulative distribution functions by upper case letters, and thus the probability that the counterpart is brighter than (i.e. magnitude smaller than) some magnitude $m$ is given by

$$C(m) = \int_{m'=-\infty}^{m'=m} c(m')\mathrm{d}m', \qquad (8)$$

which leads to

$$\frac{\mathrm{d}C}{\mathrm{d}m} = c(m). \qquad (9)$$

A distribution which is independent of crowding is the probability density function for the brightest star (be it field star or counterpart) in a given error circle, $b(m)$. This is straightforward to measure by choosing a given confidence radius and then creating a histogram of the magnitudes of the brightest stars found within each circle. However, if this is to be a useful quantity we must establish the relationship between it and $c(m)$. Since the brightest star in the error circle may be either a field star or a counterpart, we define the probability density function of the brightest field stars in our sample of error circles as $b_f(m)$. Of course both $b_f(m)$ and $b(m)$ will have their cumulative distribution functions $B_f(m)$ and $B(m)$ respectively. As in Section 3, it is important to note that $b(m)$, $b_f(m)$ and $c(m)$ are true probability distribution functions that are normalized so they integrate one. However, the answer to the question "what is the probability that the brightest star in a given X-ray circle lies between magnitudes $m$ and $m + dm$", is $Z\, b(m) dm$, where $Z$ is the probability that there is one or more stars within the error circle.

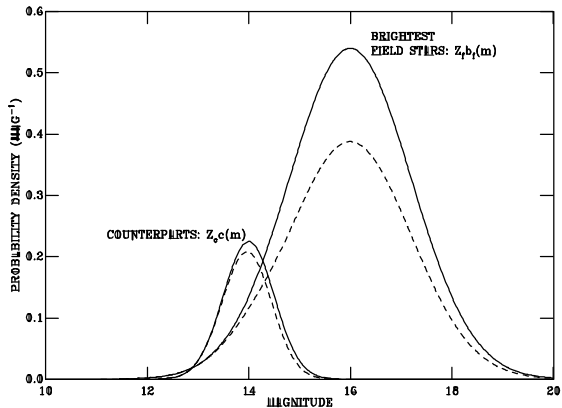

**Figure 3.** Illustrative probability distributions for the derivation of Equation 12. The probability distributions $Z_c\, c(m)$ and $Z_f\, b_f(m)$ for the counterparts and brightest field stars respectively are shown as the solid curves. The dashed curves below them show the effects of crowding. The dashed curve below $Z_c\, c(m)$ is the probability distribution for the hypothesis that the brightest star is a counterpart, and has a magnitude $m$, and is given by $Z_c\, c(m)[1 - Z_f B_f(m)]$. The dashed curve below $Z_f\, b_f(m)$ is the probability distribution for the hypothesis that the brightest star is a field star, and has a magnitude between $m$, and is given by $Z_f\, b_f(m)[1 - Z_c C]$.

Given the above, we can now consider Figure 3, which shows illustrative probability functions which might occur in a relatively crowded error circle, where it is likely the counterpart is bright, but there is a relatively large number of field stars. Consider first the probability that there is one or more field stars in the error circle *and* the brightest of those field stars has a magnitude of 16. This is given by $Z_f\, b_f(m)$ at $m = 16$, where $Z_f$ is the fraction of error circles that have field stars in them which are in our IR catalog. It is shown as the larger of the two solid curves in Figure 3. However, for this field star to be the brightest star in the error circle, there must be no counterpart brighter that outshines it. The probability there *is* a counterpart brighter than $m$ is the integral of $c(m)$ from $-\infty$ to $m$, which given our definitions above is $C(m)$, multiplied by $Z_c$, the fraction of X-ray sources that have counterparts within the error circle and in the IR catalog. This is shown as the smaller of the two solid curves in Figure 3. Thus the probability there *is not* a counterpart that outshines the field star of magnitude $m$ is $1 - Z_c\, C(m)$. This allows us to write down the probability that the brightest star in the error circle is a field star of magnitude $m$ as the combination of the following.

1. The probability there is at least one field star in the error circle; $Z_f$.

2. The probability that any field star has a magnitude $m$; $b_f(m)$.

3. The probability there is not a counterpart brighter than $m$; $[1 - Z_c\, C(m)]$.

The resulting probability distribution
$$Z_f\, b_f(m)[1 - Z_c\, C(m)], \qquad (10)$$
is shown as the dashed curve below $Z_f\, b_f(m)$ in Figure 3. This function must be symmetrical in bright stars and counterparts, and so the probability the brightest star in the error circle is a counterpart of magnitude $m$ is
$$Z_c\, c(m)[1 - Z_f\, B_f(m)]. \qquad (11)$$
In the example shown in Figure 3 the counterparts are sufficiently bright compared with the field stars that $B_f(m)$ remains small throughout the interesting range. Thus the term in brackets in Equation 11 remains close to one, ensuring that the dashed curve below $Z_f\, c(m)$ in Figure 3 remains close to the solid curve. This demonstrates that the crowding has little effect on the magnitude distribution of counterparts. By contrast the dashed line showing the effects of crowding on the field stars falls significantly below $Z_f\, b_f(m)$, showing that crowding does have an effect on these fainter stars.

We can now derive the probability distribution that there is one or more stars within the error circle, and that the brightest star is of magnitude $m$ by summing the contributions from the field stars and counterparts (Equations 10 and 11 respectively) to give
$$Z\, b(m) = Z_f\, b_f(m)[1 - Z_c\, C(m)] + Z_c\, c(m)[1 - Z_f\, B_f(m)]. \qquad (12)$$

A crucial distinction here is that the cumulative distribution function of the magnitude of the brightest field star, $B_f(m)$ is not the same as the cumulative distribution function of field-star magnitudes per unit area, $NF(m)$. The latter is what is measured if one simply plots the cumulative distribution function in magnitude for a fixed area of the field. There is a relationship between them, however, that can be derived in the following way. Let the probability density function of the magnitudes of field stars be $f(m)$, the area of the error circle $A$, and the total number of field stars (i.e. integrated over all magnitudes) per square arcsecond in the IR catalog $N$. Then the probability there is a field star between magnitudes $m$ and $m+dm$ in the circle is $AN f(m)$. The relationship between $F(m)$ and $B_f(m)$ is
$$Z_f\, B_f(m) = 1 - e^{-ANF(m)}. \qquad (13)$$
This formula has an intuitive derivation in terms of the fraction of an area covered by randomly placed objects, but is probably better derived from the Poisson distribution as the chance of not counting zero objects given a distribution with a mean of $ANf(m)$. It can be differentiated to give
$$Z_f\, b_f(m) = AN f(m) e^{-ANF(m)}. \qquad (14)$$

Combining the last three equations, and dropping the explicit statement that the probability density functions are functions of magnitude, gives
$$Z\, b = [1 - Z_c\, C] AN f e^{-ANF} + Z_c\, c\, e^{-ANF}. \qquad (15)$$

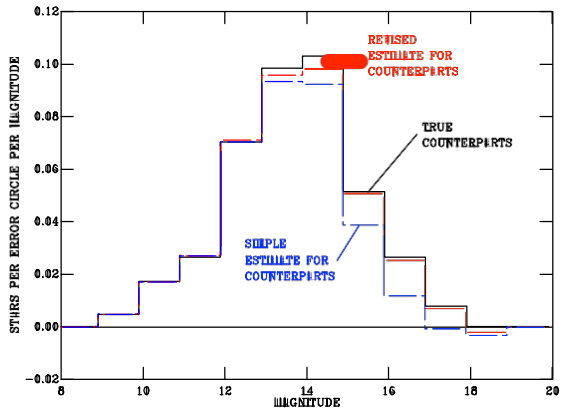

**Figure 4.** A comparison of the derived counterpart magnitude probability density functions for the Trifid Nebula field. The dashed red histogram is the result of the technique described in Section 4, the dashed blue line is the simple technique from Figure 2. The solid histogram shows the input simulated probability density function for the counterpart magnitudes.

This equation can understood as giving the probability density function of the magnitudes of the brightest stars in the error circles. The first term on the right is the contribution of the counterparts, and a second term that of the field stars. Re-arranging it to be explicit in $Z_c\, c$ gives,

$$Z_c\, c = Z\, b\, e^{ANF} - [1 - Z_c\, C]ANf. \quad (16)$$

This equation is the principal result of this section, expressing the probability density function for the distribution of counterpart magnitudes ($c$, $C$) in terms of the measurable probability density distribution for the brightest stars found in the error circles ($b$) and the probability density function for field stars ($f$, $F$). In principle this equation is for a single error circle, yet the distributions are summed over all error circles. However it is straightforward to show (by summing $c$ over many error circles) that Equation 16 holds for the same quantities averaged over all error circles.

Equation 16 can be solved by first evaluating $Z_c\, c$ and $Z\, b$ at bright values of $m$ where $C$ is zero, and then proceeding to fainter (larger) values of $m$. In Section 3 we required $Xc$, where $X$ is the fraction of X-ray sources that have counterparts anywhere in the IR catalog (not just within an error circle). This one obtains by simply dividing $Z_c\, c$ by the fraction of counterparts expected to be enclosed in the error circle radius chosen.

The improvement in the estimation of the magnitude probability density function can be seen in Figure 4. There are two reasons why the improvement in the estimate of the probability density function for the counterpart magnitudes is crucial. First if (as happens for the simple estimate between $K = 17$ and 18) the estimate of the number of sources incorrectly sinks to zero, then no counterpart stars will be found in that magnitude range. Second, this is our only way of estimating the relative number of stars with counterparts between different magnitude ranges.

## 5. APPLICATION TO THE TRIFID NEBULA FIELD

There is a series of practical issues related to using the results of Sections 3 and 4 that are best illustrated by following through a single cluster as an example. We will do this using the MYStIX Trifid Nebula field data where the X-ray sources are from the *Chandra* X-ray Observatory and the IR sources are from UKIRT data (Feigelson et al. 2013).

To test that our X-ray positions and uncertainties were reliable, all stars with $K < 15$ and within $5''$ of an X-ray position were selected and their differences in RA and Dec from the X-ray position calculated. This distribution was then fitted as a uniform field-star distribution plus a two-dimensional Gaussian distribution whose width was that expected from the X-ray data analysis. When the central position of the Gaussian was allowed to run free in $\delta$RA and $\delta$Dec we found a global shift of $+0.1''$ in both axes. After correcting this offset, the Gaussian width was allowed to be a free parameter, parameterized as a scale factor times the given error circle radius, with a further fixed value ("systematic") uncertainty added in quadrature. The best fitting scale factor was 0.78, which is very close to the $1/\sqrt{2}$ one might expect since the X-ray positional uncertainty from Kuhn et al. (2013a) is given as $\sigma_{63}$ in the formula $e^{(-r^2/\sigma_{63}^2)}$, compared with the definition used here of $e^{-0.5(r^2/\sigma_{39}^2)}$ (see Appendix B). Thus the scale factor was fixed at $1/\sqrt{2}$ which yielded a systematic uncertainty of $0.072''$. Since the systematic uncertainty is independent of the accuracy of the X-ray position, we believe it represents the uncertainty in position in the IR catalog. Its value is very close to the radius which includes 68% of stars from our IR astrometric solution, which is approximately $0.13''$ (King et al. 2013), which translates to $0.086''$ in our definition of $\sigma_{39}$.

To derive the counterpart magnitude probability density function, $c(m)$, as described in Section 4 we must first decide an area of error circle from which to draw the data to construct the histogram. This problem is exactly analogous to the one encountered in aperture photometry, where the issue is what radius will optimize the signal-to-noise. There is no one radius which is optimal over all ratios of background to object counts, but a defensible choice is to optimise for the case when there are many more field stars than counterparts, since this is where it is crucial we obtain the best model. Naylor (1998) showed that in this background-limited case the optimal radius is 0.6 times the full width at half maximum, which corresponds to $1.4\sigma_{39}$. We actually chose to use $1.51\sigma_{39}$ since this encloses 68% of all X-ray counterparts. The field-star contamination is estimated using an annular region whose inner radius is equal to that within which 99% of X-ray counterparts will fall, and an outer radius of twice that.

There are two pathological cases of $c(m)$ that are caused by low number statistics. Our first provision against small-number statistics is to determine $c(m)$ from a large number of error circles. We used the nearest 400 positions to the position in question, and bins of one magnitude. Even then there were some positions where inaccuracy in the subtraction of the field star magnitude distribution led to a negative $c(m)$. These occur mostly at the faint end of the distribution and we replaced them with zero. There were similar problems at the bright end of the distribution where both $c(m)$ and the field star magnitude probability density function $f(m)$ became zero. This could lead to very bright stars being unassigned, when actually the correlation of an X-ray position with a bright star is so unusual that it should almost certainly be assigned as a counterpart. This led us to extrapolate both $c(m)$ and $f(m)$ to arbitrarily bright magnitudes, using $c(m) \propto 3^m$ as a simple analytical function that represented the data at somewhat fainter (larger) magnitudes. In fact, the precise form is not important, as long as the ratio $c(m)/f(m)$ is large at bright $m$.

Having established the counterpart magnitude probability density function, we then needed to evaluate Equations 6 and 7 for each star in each error circle. This requires a much more precise determination of the field-star magnitude probability density function for than the one used for determining the overall counterpart magnitude distribution, since $Nf(m)$ is required separately for each error circle. Therefore it was determined using an inner radius of $5''$, and an initial outer radius of $50''$. If this resulted in fewer stars defining the field-star probability density function than had defined the counterpart distribution, the radius was increased until this criterion was passed. The distribution of counterparts with position $g(\Delta x_i, \Delta y_i)$ is simply a two dimensional Gaussian normalized to integrate to one, and $Xc(m)$ is derived as described in Section 4. We used all stars out to a radius of either the 99.9% confidence radius, or the radius at which $g(\Delta x_i, \Delta y_i) = 0.1Nf$, whichever was the greater.

The final output from this process is a probability for each star that it is the counterpart, and the probability that there is no counterpart. These data are provided only as electronic tables, with Table 2 (which appears in the printed edition) giving the column headings. Electronic Tables 3 and 4 give the NIR and *Spitzer* matches respectively for all MYStIX fields with these datasets listed in Table 2 of Feigelson et al. (2013). The tables give all stars which have a likelihood of being the counterpart of greater than 0.05 (assuming a star falls within our search radius defined above), the probability that each such is the counterpart, and the probability there is no counterpart in the IR catalog. The number of sources in each waveband, and the numbers of matches obtained for each MYStIX field are given in Table 1 of Broos et al. (2013).

The most straightforward way of using these data is accept the most likely "model", whether it be that a given star is the counterpart, or that there is no observed counterpart. However, there are examples where the statistics imply that no definitive conclusion can be reached, for example when the probability for the most likely counterpart falls below 0.5. We therefore accept as counterparts for input into the Bayesian classifier (Broos et al. 2013) only objects which have a probability of greater than 0.8 of being the correct counterpart. Sources that satisfy this probability threshold as being a counterpart to an X-ray source are assigned MATCH_FLG = 1.

## 6. COMPARISON WITH NEAREST NEIGHBOR MATCHING

A useful benchmark is the comparison of our matching results to those from proximity-only matching. Our X-ray catalog for the Trifid Nebula field contains 633 sources; the Bayesian method finds counterparts for 395 of these (at a probability of more than 0.8) and that there is no counterpart in the NIR catalog for 137 of them (again with a probability of more than 0.8). For comparison we define a proximity-only sample as consisting of the closest star to each X-ray position, provided it falls within the 99% confidence radius, which yields 507 counterparts. The difference between the two samples is illustrated in Figure 5, where for each counterpart its $K$-band magnitude as plotted as a function of distance from the X-ray position, normalised by the uncertainty in X-ray position. The stars that have a probability derived by our Bayesian method of greater than 0.99 of being the counterpart are marked with blue asterisks, and those whose probability lies between 0.99 and 0.80 with red crosses. The borderline at a probability of 0.99 has an obvious negative slope (recall that the magnitude axis is reversed). This reflects the fact that faint stars far from the X-ray position are more likely to be field stars that stars of similar magnitude close to the position.

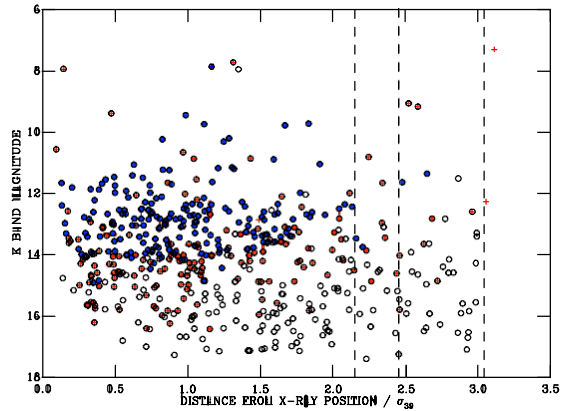

**Figure 5.** A comparison of the properties of the stars chosen as counterparts in the Trifid Nebula field field by proximity-only matching and our Bayesian technique. The symbols show Bayesian matches with probabilities of being the counterpart of greater than 0.99 (blue asterisks) and between 0.99 and 0.8 (red crosses). Proximity-only matches are shown as open circles. The y-axis is $K$-band magnitude of the star, the x-axis its distance from the X-ray position in units of the error circle radius (see Appendix B). The dotted lines represent the edges of circles that would contain 90, 95 and 99% of the counterparts, and so would be reasonable positions to cut off a closest match search.

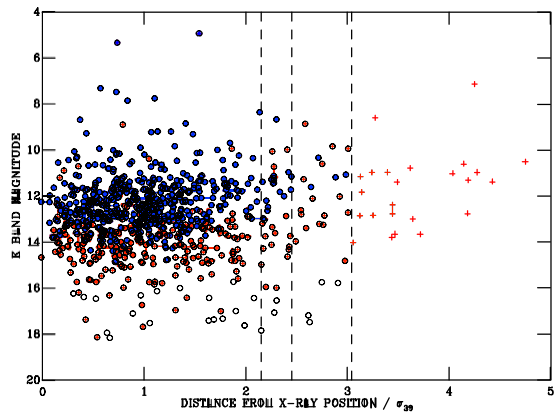

**Figure 6.** As Figure 5 but for NGC2264, and with the division between asterisks and crosses being drawn at 0.999.

The open black circles in Figure 5 mark the proximity-

Table 2
Matching results

| Column Label (Table 3) | Column Label (Table 4) | Units | Description |
| --- | --- | --- | --- |
| XRAY_NAME | XRAY_NAME | | X-ray catalog source name |
| XN_PROB_NO_CT | XM_PROB_NO_CT | | Probability that no counterpart exists in the IR catalog |
| XN_PROB_CP | XM_PROB_CP | | Probability for the first match |
| RA | RA | deg | Right ascension (J2000) for first match |
| DEC | DEC | deg | Declination (J2000) for first match |
| XN_PROB_CP_2 | XM_PROB_CP_2 | | Probability for the second match |
| RA_2 | RA_2 | deg | Right ascension (J2000) for second match |
| DEC_2 | DEC_2 | deg | Declination (J2000) for second match |
| XN_PROB_CP_3 | XM_PROB_CP_3 | | Probability for the third match |
| RA_3 | RA_3 | deg | Right ascension (J2000) for third match |
| DEC_3 | DEC_3 | deg | Declination (J2000) for third match |
| XN_PROB_CP_4 | | | Probability for the fourth match |
| RA_4 | | deg | Right ascension (J2000) for fourth match |
| DEC_4 | | deg | Declination (J2000) for fourth match |

**Note.** — Columns are only filled if there are counterparts with $P(i) > 0.05$.

only counterparts. Where circles are filled with crosses or asterisks it shows the proximity-only and Bayesian methods agree as to which star is the counterpart, but it is clear there is a large number (almost 100) of faint proximity-only counterparts that the Bayesian method rejects. This is the behavor we expect; the Trifid Nebula field is so crowded in the NIR that it is quite likely that a faint star within an error circle is not the counterpart. These "empty circles" are not simply cases of stars which have a significant probability of being counterparts, but do not reach our 0.8 threshold. For around half of these positions the Bayesian method gives a probability of more than 0.8 that there is no counterpart in the NIR catalog, and only two of these faint stars chosen only by the proximity method exceed a probability of 0.5 of being the counterpart. This means that roughly 20% of the counterparts identified by the proximity-only method are probably incorrect. Thus we conclude that in crowded fields the Bayesian procedure outperforms proximity-only matching, specifically by rejecting as counterparts faint field stars that would be erroneously included in a catalog derived from taking the closest neighbors.

One would expect there to be cases where there is a faint star close to the X-ray position, and a brighter one a little further away, which causes the Bayesian method to chose the brighter star, whilst the proximity-only method choses the fainter. In fact there are only five such cases in our Trifid Nebula field catalog, and for three of those the brighter star does not reach the probability requirement of 0.8 to be taken as a counterpart. So, whilst the displacement or problem outlined in Section 1 is good example of how the proximity-only matching might go wrong, the major displacement problem is that faint stars in the catalogue are chosen as counterparts when the true counterpart probably lies below the completeness limit of the IR data. In the terminology of Broos et al. (2011) (see their Figure 4), it is not the "false matches" which the Bayesian technique suppresses, it is the "false positives".

In a more practical proximity-only matching scheme one could limit the number of false positives by choosing a smaller matching radius. This is illustrated in Figure 6 of Feigelson et al. (2013) where the rate of false positives and true matches are plotted as a function of matching radius. It makes clear that for the Trifid Nebula field a smaller radius than 99% would be appropriate.

Figure 6 of this paper shows the same plot as Figure 5 but for the NGC2264 catalog, which is a much less crowded in the NIR than the Trifid Nebula field. The divide between the asterisks and the crosses in Figure 6 is set at a probability of 0.999, rather than the 0.99 used for the Trifid Nebula field. These numbers were chosen to divide the sample of stars with counterpart probabilities greater than 0.8 roughly in half, and so this change tells us that in the less crowded field, counterparts are, on the whole, more likely to be correct. There is still a small number of faint objects that are chosen by the proximity-only matching, which are rejected by our technique. The decline in the number of these is simply that in a less crowded field, it is less likely that proximity-only matching will find an unrelated star within the error circle.

In the NGC 2264 field there are 24 objects that the Bayesian procedure identifies as possible counterparts with a probability of between 0.999 and 0.8 that lie beyond the 99% confidence radius. Given there are a total of 763 matches, we would expect only eight outside this radius. This population represents just 4% of the sources brighter than $K = 14$, and it is possible that our astrometric uncertainties are non-Gaussian at this level. Conversely, the median offsets between the X-ray and IR positions for this subgroup is 0.8″[11], well below the resolution of *Chandra* and just two pixels (with typical seeing of 2.5 pixels) of the UKIRT data. Hence astrophysical explanations, such as fainter stars lying undetected close to bright objects are also possible. This interpretation is supported by the results of cross-matching these 24 objects with the optical catalog of Mayne et al. (2007), which has usable optical photometry for 14 of the sample. Of those 12 lie in the pre-main-sequence region of the CMD. Thus it seems likely that a significant fraction of these stars are in pre-main-sequence binary systems where one of the objects is an X-ray source.

7. COMPLETENESS

In the ideal case of perfect X-ray and IR positions in an uncrowded field, the counterpart list would contain all pairings of X-ray and IR sources whose fluxes lay above some specified limits in both wavebands. Of course, in any band there is no hard flux limit, instead there is a completeness function. So the completeness of the counterparts list at a given X-ray and IR flux is the product of the individual completeness functions for each catalog, evaluated at those fluxes. Imperfect positions and crowding add the possibilities of false positives, false negatives and incorrect matches, whose occurrence will primarily be a function of IR magnitude. If a sample is chosen with a very high $P(i)$ then false positives and incorrect matches will have little impact upon it, and only false negatives will be important. The fraction of false negatives can be estimated by dividing the distribution of magnitudes of accepted counterparts by the true distribution derived in Section 4 and shown in Figure 4. Hence the counterpart list completeness at a given X-ray and IR flux is the completeness of the two catalogs multiplied by the fraction of false negatives at the appropriate IR magnitude.

## 8. AN APPROACH FOR MATCHING MANY CATALOGS

In Section 1 we introduced what appears to be a fundamental asymmetry between our catalogs by introducing the terms "master" for the X-ray catalog and "slave" for the IR data. In fact, the equations remain symmetric provided one uses only differences in position, since they are independent of which catalog is considered first. The asymmetry is introduced in Section 3 when one considers the probability that the counterpart will have an IR magnitude that is drawn from a different distribution from the IR magnitude distribution of field stars. Specifically it is the term $Xc(m)/Nf(m)$ which determines which is the master and which the slave catalog. Clearly the symmetry can be regained by introducing a similar term that compares the likelihood that an X-ray source with an IR counterpart has a given X-ray flux with the likelihood that an X-ray source without an IR counterpart has that flux.

With that we believe all the pieces are now in place for a full solution to the problem of matching sources from many different wavebands to give spectral energy distributions for those sources. One could construct the network of probabilities that a given source is related to sources in other bands, and then propose hypotheses that relate particular sources across bands, and find the most likely.

Such a procedure would differ from that of Budavári & Szalay (2008) in that one does not hypothesize a particular model for the spectral energy distribution, which has the advantage of avoiding using astrophysical information. It is closer to the approach of Storkey et al. (2005), although they determine the distribution of magnitudes for sources that are counterparts at the same time as determining which sources are related. The best solution is then iterated using the expectation maximization algorithm, whereas our solution of determining it beforehand, whilst potentially less accurate, is probably good enough and both computationally and conceptually more straightforward.

## 9. CONCLUSIONS

We have shown that in typical Galactic plane survey data from the new generation of NIR surveys such as UKIDSS (Lucas et al. 2008; Lawrence et al. 2007) and those from VISTA (Emerson et al. 2004), choosing the closest NIR star to an X-ray position can result in a significant fraction of the NIR counterparts being incorrectly identified. This is because the density of faint stars is so high that many X-ray error circles will, by chance, contain a faint star that is unrelated to the X-ray source. We have presented a method for finding reliable matches between X-ray and NIR sources in the face of this extreme NIR crowding. The method utilizes the data themselves to determine the magnitude distributions of both the field stars and the counterparts to the X-ray sources (see Section 4). This is then used in the Bayesian framework of Sutherland & Saunders (1992) to derive a probability that any given star is the counterpart (Section 3).

Applying this method to the case of the Trifid Nebula field, where the crowding is particularly dense, we show that roughly 20% of the counterparts are mis-identified by classical matching. In four MYStIX fields (the Trifid Nebula, the Eagle Nebula, NGC6357 and M17) more than 10% of stars which are the closest objects to an X-ray position and within a 99% confidence error circle are identified as field stars by our technique, and that those stars are concentrated systemically at faint magnitudes (Section 6).

## ACKNOWLEDGEMENTS

The MYStIX project is supported at Penn State by NASA grant NNX09AC74G, NSF grant AST-0908038, and the Chandra ACIS Team contract SV4-74018 (G. Garmire & L. Townsley, Principal Investigators), issued by the Chandra X-ray Center, which is operated by the Smithsonian Astrophysical Observatory for and on behalf of NASA under contract NAS8-03060. This research made use of data products from the *Chandra* Data Archive and the *Spitzer* Space Telescope, which is operated by the Jet Propulsion Laboratory (California Institute of Technology) under a contract with NASA.

This work is based in part on data obtained in Director's Discretionary Time on the UK Infrared Telescope, and in part on data from the UKIRT Infrared Deep Sky Survey. The United Kingdom Infrared Telescope is operated by the Joint Astronomy Centre on behalf of the Science and Technology Facilities Council of the U.K.

This publication makes use of data products from the Two Micron All Sky Survey, which is a joint project of the University of Massachusetts and the Infrared Processing and Analysis Center/California Institute of Technology, funded by the National Aeronautics and Space Administration and the National Science Foundation.

## REFERENCES

Broos, P. S., Townsley, L. K., Feigelson, E. D., et al. 2011, ApJS, 194, 2
Broos, P. S., Getman, K. V., Povich, M. S., et al. 2013, ApJS, submitted [MYStIX MPCM paper]
Brusa, M., Zamorani, G., Comastri, A., et al. 2007, ApJS, 172, 353
Budavári, T., & Szalay, A. S. 2008, ApJ, 679, 301
Emerson, J. P., Sutherland, W. J., McPherson, A. M., et al. 2004, The Messenger, 117, 27
Feigelson, E. D., Getman, K., Townsley, L., et al. 2005, ApJS, 160, 379
Feigelson, E. D., Townsley, L. K., Broos, P. S., et al. 2013, ApJS, submitted [MYStIX overview paper]
Fleuren, S., Sutherland, W., Dunne, L., et al. 2012, MNRAS, 423, 2407

King, R. R., Naylor, T., Broos, P. S., Getman, K. V., & Feigelson, E. D. 2013, ApJS, submitted [MYStIX X-ray matching paper]

Kuhn, M. A., Getman, K. V., Broos, P. S., Townsley, L. K., & Townsley, L. K. 2013a, ApJS, submitted [MYStIX 1st X-ray catalog paper]

Kuhn, M. A., Povich, M. S., Luhman, K. L. Getman, K. V., Busk, H. S., & Feigelson, E. D. 2013b, [MYStIX mid-IR catalog paper]

Lawrence, A., Warren, S. J., Almaini, O., et al. 2007, MNRAS, 379, 1599

Lucas, P. W., Hoare, M. G., Longmore, A., et al. 2008, MNRAS, 391, 136

Luo, B., Brandt, W. N., Xue, Y. Q., et al. 2010, ApJS, 187, 560

Mann, R. G., Oliver, S. J., Serjeant, S. B. G., et al. 1997, MNRAS, 289, 482

Mayne, N. J., Naylor, T., Littlefair, S. P., Saunders, E. S., & Jeffries, R. D. 2007, MNRAS, 375, 1220

Naylor, T. 1998, MNRAS, 296, 339

Oyabu, S., Yun, M. S., Murayama, T., et al. 2005, AJ, 130, 2019

Povich, M. S.and Kuhn, M. A., Getman, K. V., Feigelson, E. D., et al. 2013, [MYStIX infrared excess paper]

Rohde, D. J., Drinkwater, M. J., Gallagher, M. R., Downs, T., & Doyle, M. T. 2005, MNRAS, 360, 69

Rohde, D. J., Gallagher, M. R., Drinkwater, M. J., & Pimbblet, K. A. 2006, MNRAS, 369, 2

Roseboom, I. G., Oliver, S., Parkinson, D., & Vaccari, M. 2009, MNRAS, 400, 1062

Rutledge, R. E., Brunner, R. J., Prince, T. A., & Lonsdale, C. 2000, ApJS, 131, 335

Storkey, A., Williams, C., Taylor, E., & Mann, R. G. 2005, http://www.anc.ed.ac.uk/~amos/publications/StorkeyEtAl2005AnEMAlgorithmForRecordLinkage.pdf, Tech. rep., University of Edinburgh

Sutherland, W., & Saunders, W. 1992, MNRAS, 259, 413

Telleschi, A., Güdel, M., Briggs, K. R., Audard, M., & Palla, F. 2007, A&A, 468, 425

Townsley, L. K., & Broos, P. S. 2013, ApJS, submitted [MYStIX 2nd X-ray catalog paper]

## APPENDIX

## A DERIVATION FROM BAYES' RULE

The proof given for Equations 6 and 7 in Section 3 is derived from the fundamentals of the problem, and does not use Bayes' rule directly. A direct derivation from Bayes' Rule is perhaps more rigorous, and certainly establishes the position of these equations within modern statistical theory, though perhaps gives less insight into the problem. For such a proof, we begin by considering two hypotheses. The first is that there is no counterpart in the catalog, $H_0$, and the second that there is a counterpart in the catalog $\tilde{H}_0$, i.e. not $H_0$.

To apply Bayes' Rule requires the likelihoods that $H_0$ and $\tilde{H}_0$ will result in the data, $L(D|H_0)$ and $L(D|\tilde{H}_0)$. These can be derived by first considering a simplified case where all stars have the same magnitude, and the stars are placed in a grid of infinitesimal pixels. The likelihoods for this case we shall denote with primes, $L'(D|H_0)$ and $L'(D|\tilde{H}_0)$. The pixels are of area $dy\,dx$ and each of them has a probability of being occupied by a field star of $O$ (the occupancy). $O$ is so small there is no chance of double occupancy. For a dataset with $M$ of the pixels occupied, $L'(D|H_0)$ is the likelihood that a model consisting only of field stars will produce the observed pattern of filled and empty pixels. It can be evaluated by working sequentially through each pixel, and if it is occupied in the dataset, the model has a likelihood $O$ of matching it, if not $1-O$. Thus the likelihood of the model where there is no counterpart in the data matching the data pattern is $L'(D|H_0) = O^M \times (1-O)^{T-M}$ where $T$ is the total number of pixels.

Now consider the model where there is a counterpart to the X-ray source in the dataset. This model can only produce the observed catalog if the counterpart is one of the stars, and the other $M-1$ stars originate from the field. Cycling the counterpart through all $M$ stars gives the $M$ ways $\tilde{H}_0$ can produce the data. Consider the case where star $j$ is the X-ray source. This means the field must produce the observed pattern less star $j$, i.e. a pattern of $M-1$ stars, for which the likelihood is $O^{M-1} \times (1-O)^{T-M+1} = L'(D|H_0)(1-O)/O \approx L'(D|H_0)/O$, since $O$ is small. This must be multiplied by the probability that the counterpart has fallen in whichever pixel star $j$ occupies. This is simply $g(\Delta x_j, \Delta y_j)\,dy\,dx$. So the likelihood that a model where star $j$ is the counterpart will produce the observed catalog is

$$L'(D|H_j) = L'(D|H_0) \frac{g(\Delta x_j, \Delta y_j)\,dy\,dx}{O}. \tag{A1}$$

The sum over all $j$ then gives

$$L'(D|\tilde{H}_0) = \sum_j L'(D|H_j) = L'(D|H_0) \sum_j \frac{g(\Delta x_j, \Delta y_j)\,dy\,dx}{O}. \tag{A2}$$

The next step is to remove the infinitesimals by expressing $O$ in terms of other quantities. The probability any pixel is occupied is approximately the number of stars per unit area in the catalog, $N$, multiplied by the area of a pixel, $dy\,dx$. Hence $O = N\,dy\,dx$, and so our expression for the likelihood of $\tilde{H}_0$ is

$$L'(D|\tilde{H}_0) = L'(D|H_0) \sum_j \frac{g(\Delta x_j, \Delta y_j)}{N}. \tag{A3}$$

We can now change from $L'$ to $L$ by multiplying $L'$ by the probability that the stars lie at the observed magnitudes. For $L(D|H_0)$ the factor is $\prod_i f(m_i)$ where the product is taken over all stars in the catalogue. Thus

$$L(D|H_0) = L'(D|H_0) \prod_i f(m_i). \tag{A4}$$

**Table 5**
Ratios of error circle radii. The results are the numbers obtained if the radius of the error circle that encloses the percentage given on the top row is divided by the radius given in the first column.

|      | 99.0 | 95.0 | 90.0 | 68.2 | 63.2 |
|------|------|------|------|------|------|
| 39.3 | 3.04 | 2.45 | 2.15 | 1.51 | 2    |
| 63.2 | 2.15 | 1.73 | 1.52 | 1.07 |      |
| 68.2 | 2.00 | 1.62 | 1.42 |      |      |
| 90.0 | 2    | 1.14 |      |      |      |
| 95.0 | 1.24 |      |      |      |      |

For each term in Equation A3 which contributes to $L(D|\tilde{H}_0)$ there will be a similar factor, except the sum must omit star $j$, and include the probability that the counterpart has a magnitude $m_j$, $c(m_j)$, hence

$$L(D|\tilde{H}_0) = L'(D|H_0) \sum_j \frac{c(m_j)g(\Delta x_j, \Delta y_j)}{N} \prod_{i \neq j}^{n} f(m_i) = \sum_j \frac{c(m_j)g(\Delta x_j, \Delta y_j)L(D|H_0)}{Nf(m_j)}. \quad (A5)$$

The remaining terms in Bayes' Rule

$$P(H_0|D) = \frac{P(H_0)L(D|H_0)}{P(H_0)L(D|H_0) + P(\tilde{H}_0)L(D|\tilde{H}_0)}, \quad (A6)$$

are the priors, which are simply the fraction of stars which do not have counterparts

$$P(H_0) = 1 - X \quad (A7)$$

and the fraction which do

$$P(\tilde{H}_0) = X. \quad (A8)$$

This yields

$$P(H_0|D) = \frac{1 - X}{1 - X + \sum_j \frac{Xc(m_j)g(\Delta x_j, \Delta y_j)}{Nf(m_j)}} \quad (A9)$$

and

$$P(\tilde{H}_0|D) = \frac{\sum_j \frac{Xc(m_j)g(\Delta x_j, \Delta y_j)}{Nf(m_j)}}{1 - X + \sum_j \frac{Xc(m_j)g(\Delta x_j, \Delta y_j)}{Nf(m_j)}}. \quad (A10)$$

If we now consider the hypothesis $H_i$, that star $i$ is the counterpart, as we have no prior information as to which star is the counterpart

$$P(H_i|D) = \sum_i P(H_i|D) \frac{L(D|H_i)}{\sum_i L(D|H_i)} = P(\tilde{H}_0|D) \frac{L(D|H_i)}{L(D|\tilde{H}_0)} = \frac{\frac{Xc(m_i)g(\Delta x_i, \Delta y_i)}{Nf(m_i)}}{1 - X + \sum_j \frac{Xc(m_j)g(\Delta x_j, \Delta y_j)}{Nf(m_j)}}. \quad (A11)$$

Equations A9 and A11 correspond to Equations 6 and 7 derived in Section 3.

### THE DEFINITIONS OF $\sigma$

For a two-dimensional Gaussian there are at least three definitions of the quantity $\sigma$ in common usage, and so to avoid confusion in this paper we refer to them by using a subscript that represents the percentage of the probability a circle of radius $\sigma$ encloses. Sometimes authors will quote an error circle that encloses 68.2% of the probability (a definition of $1\sigma$ in one dimension) which we denote $\sigma_{68}$. However, it is perhaps more logical to take the definition of $\sigma$ from the two-dimensional formulae $e^{-(r/\sigma_{63})^2}$ or $e^{-0.5(r/\sigma_{39})^2}$, where $\sigma_{63}$ and $\sigma_{39}$ enclose 63.2 and 39.3% of the probability respectively. If one works with independent Gaussians in $x$ and $y$ of the form $e^{-0.5(x/\sigma_x)^2 - 0.5(y/\sigma_y)^2}$ then if $\sigma_x = \sigma_y$ $e^{-0.5(x^2+y^2)/\sigma_x^2} = e^{-0.5(r/\sigma_x)^2}$, and thus $\sigma_x = \sigma_{39}$. The relationships between all these radii are given in Table 5.